\documentclass[aps,pra,twocolumn,superscriptaddress,10pt]{revtex4-2}

\usepackage[utf8]{inputenc}
\usepackage[T1]{fontenc}
\usepackage{graphicx}
\usepackage{amsmath,amssymb,amsfonts}
\usepackage{algorithmic}
\usepackage{hyperref}
\usepackage{booktabs}
\usepackage{multirow}
\usepackage{tabularx}
\usepackage{xcolor}

\usepackage{tikz}
\usetikzlibrary{arrows.meta, positioning, shapes, calc}
\definecolor{myCyanFill}{RGB}{224,255,255}
\definecolor{myCyanDraw}{RGB}{0,204,204}
\definecolor{myOrangeFill}{RGB}{255,229,204}
\definecolor{myOrangeDraw}{RGB}{255,163,0}
\definecolor{myRed}{RGB}{204,0,0}

\begin{document}

\title{Long-term Performance Analysis of a Commercial QKD Device Under Tropical Climate Conditions}

\author{Alisson Tezzin}
\email{alisson.cordeiro@ime.eb.br}
\affiliation{Military Institute of Engineering (IME), Postgraduate Program in Defense Engineering, Rio de Janeiro, RJ, Brazil}
\affiliation{Embrapii Competence Center in Quantum Technologies (QuIIN), SENAI CIMATEC, Salvador, Bahia, Brazil}

\author{Gustavo M. Uhdre}
\affiliation{Military Institute of Engineering (IME), Postgraduate Program in Defense Engineering, Rio de Janeiro, RJ, Brazil}

\author{Oscar Martins}
\affiliation{Military Institute of Engineering (IME), Department of Computer Engineering, Rio de Janeiro, RJ, Brazil}

\author{Sabrina Rufo}
\affiliation{Institute of Computing (IComp), Federal University of Amazonas, Manaus, Brazil}

\author{Vitor G.A. Carneiro}
\affiliation{Military Institute of Engineering (IME), Postgraduate Program in Defense Engineering, Rio de Janeiro, RJ, Brazil}

\begin{abstract}
Quantum key distribution (QKD) has reached a commercially viable stage, with several companies offering hardware systems designed for operational deployment. Evaluating the performance of commercial QKD devices under real-world deployment conditions is essential for users seeking to understand the practical limitations and operational reliability of these systems. In this paper, we present a long-term performance analysis of ID Quantique's Clavis XGR deployed within the Hermes Quantum Network, in Brazil. Our study provides a detailed characterization of key operational metrics, such as secret key rate, quantum bit error rate (QBER), visibility, and detection counts, mapping their behavior over extended periods of continuous operation. We analyze the system's stability across two distinct optical links: a 40.3~km indoor spooled fiber and a 3.5~km outdoor deployed underground fiber. Monitored under both unregulated tropical ambient fluctuations and actively controlled thermal stress, our results demonstrate excellent overall baseline resilience, with the system maintaining visibility above $97\%$ and QBER below $1\%$ on average. These findings provide practical insights into the expected behavior and thermal bottlenecks of commercial QKD systems in field deployments, particularly in tropical climates, helping to inform realistic expectations for operational quantum-safe infrastructures.
\end{abstract}

\maketitle

\section{Introduction} \label{sec:introduction}  

Quantum Key Distribution (QKD) has transitioned from a theoretical cryptographic concept into a commercially viable technology, driving the establishment of secure communication networks worldwide \cite{Pirandola2020advancesQuantumCryptography,Stanley2022ProgressQKDNetworks,Zhang2025towardsGlobalQKD,bozzio2025beyondQKD,rehman2026companies}. By guaranteeing security through the fundamental laws of quantum mechanics, QKD protocols, such as the seminal BB84 \cite{bennett2014quantum}, ensure that any eavesdropping attempt introduces detectable disturbances. As public and private sectors increasingly adopt QKD to future-proof data against emerging quantum threats, addressing the practical challenges of real-world deployments has become fundamental for establishing reliable quantum-safe infrastructures.

In fiber-based field environments, environmental fluctuations---most notably temperature variations---constitute a primary bottleneck for continuous and stable key generation. Thermal stress impacts both the optical transmission channel and the QKD hardware itself. In deployed fibers, temperature shifts alter propagation delay, birefringence, and polarization states, which directly degrade interferometric visibility in time-bin or phase-encoded systems \cite{Agrawal2010fiberOpticSystems,ma2021thermalBirefringence}. On the hardware side, ambient heat exacerbates dark count rates and afterpulsing in photodetectors \cite{Sze2006semiconductorDevices,hofbauer2018temperatureDarkCountSinglePhoton,Hadfield2009singlePhotonDetectors}.

A realistic assessment of QKD reliability thus requires long-term evaluation under authentic environmental stress. To date, the vast majority of field deployments and characterizations of commercial QKD systems have been conducted in the temperate or subtropical climates of Europe, North America, and East Asia, whereas performance data reflecting the operational dynamics in tropical climates---characterized by high heat and significant thermal variations---are relatively scarce. 

To provide these practical insights, this paper presents a comprehensive long-term performance analysis of a commercial QKD device---the ID Quantique Clavis XGR---deployed within the Hermes Quantum Network, an initiative led by the Military Institute of Engineering (IME) aimed at establishing a secure quantum communication infrastructure in Rio de Janeiro, Brazil \cite{moreira2024nationalSovereignty,zanetti2025brazilianQKDspace}. We systematically evaluate the system's stability across two distinct optical topologies: an indoor channel comprising 40.3 km of spooled fiber and a 3.5 km outdoor deployed underground fiber connecting IME to the Brazilian Center for Research in Physics (CBPF).


Our experimental campaign is divided into two distinct phases. In the first phase, we map the system's autonomous response to routine thermal cycling through continuous, long-term monitoring under unregulated baseline cooling. In the second phase, we actively control the laboratory temperature to isolate the thermal effects on the hardware, thereby mitigating confounding parameters that could influence the observed correlations. By subjecting the infrastructure to the intrinsic stresses of a tropical deployment, we establish a detailed operational baseline that informs realistic expectations for quantum-safe networks, especially under tropical climate conditions.

This paper is organized as follows. Section \ref{sec:experimentalSetup} briefly presents the Clavis XGR system architecture and the optical network topology. Section \ref{sec:experimentalProcedure} explains the experimental procedure, detailing the two experimental phases. Section \ref{sec:results} presents our results, which are discussed in Section \ref{sec:discussion}.

\section{System Architecture and Experimental Setup}\label{sec:experimentalSetup}

The experiments reported in this work were performed using an ID Quantique Clavis XGR system, which is a modular research platform for fiber-based quantum key distribution implementing a biased BB84 \cite{lo2005efficientQKDandProof} with time-bin \cite{brendel1999pulsed,tittel2000CryptographyEnergyTimeBell} and decoy state protocol \cite{lo2005decoy,hwang2003quantum}. The system follows a standard Alice–Bob architecture, in which the transmitter (Alice) prepares phase-modulated optical pulses and the receiver (Bob) performs interferometric measurements for quantum state decoding. 
\subsection{Clavis XGR Architecture and biased BB84 Protocol}\label{subsec:clavisArchitecture}

As the complete proprietary hardware architecture of the Clavis XGR is not publicly disclosed, the following description is synthesized from a public technical brochure \cite{IDQ2025clavisBruchure} and the operational mechanisms inferred through the system's accessible research functionalities.

At the transmitter (Alice), a 1 GHz pulsed weak coherent laser operating at a wavelength of 1551.72 nm generates optical pulses that pass through an unbalanced Mach-Zehnder interferometer. This splits each pulse into a pair of coherent pulses propagating in the same spatial mode but separated by a $\leq 1$ ns time-bin, defining distinct early and late states. To encode information in the computational Z-basis, an intensity modulator selectively extinguishes either the first or second pulse. For the  X-basis, a phase modulator introduces a relative phase difference of $\varphi=0$ or $\varphi=\pi$ (in or out of phase) between the consecutive pulses to generate the required superposition states. All stochastic decisions are governed by an integrated Quantum Random Number Generator (QRNG). Finally, the prepared optical qubit is transmitted to the receiver (Bob) through the quantum channel.

Unlike the canonical BB84 scheme, a biased basis choice is made to optimize the secret key rate \cite{lo2005efficientQKDandProof,kawakami2025securityBB84passive}. For this study, the protocol utilized the system's default asymmetric probability distribution, selecting the Z-basis with a 75\% probability and the X-basis with a 25\% probability. In this configuration, the highly probable Z-basis is used for secure key generation, whereas the X-basis serves to monitor channel disturbances and tightly bound the eavesdropper's information \cite{kawakami2025securityBB84passive}. 

To implement the decoy-state method, the intensity modulator subsequently attenuates the prepared states into three different intensity levels. We maintained the platform's default, yet configurable, intensity and probability parameters: a signal state ($\mu_0 = 0.45$) with 20\% probability, a weak decoy state ($\mu_1 = \mu_{0}/2 = 0.225$) with 60\% probability, and a vacuum state ($\mu_2 = 0$) with 20\% probability, in which $\mu_i$ is the average number of photons per qubit of each state.

The receiver module (Bob) utilizes internal single-photon detectors. Decoding is performed via a passive basis choice: an optical beam splitter routes the incoming quantum signals into two measurement paths. In the Z-basis path, a single detector directly records the photon's time of arrival to distinguish between the early and late time-bins. Conversely, the X-basis path routes the signal through an unbalanced Mach-Zehnder interferometer---matching the $\leq 1$ ns delay of Alice's internal interferometer---to interfere the consecutive pulses. The two output ports of this interferometer are coupled to two separate detectors, enabling the discrimination of the relative phase ($\varphi=0$ or $\pi$) of the superposition states. Rounds presenting coincident clicks in distinct detectors (whether within the X-basis or across both measurement paths) are discarded, leaving only unambiguous single-detection events to proceed to the sifting phase. To support secure classical post-processing and local randomness requirements, Bob is also equipped with an integrated QRNG.

The overall architecture relies on a multi-channel network topology. Alongside the unidirectional quantum channel used for qubit transmission, the system has a bidirectional service channel---operating via standard C-band transceivers over a dedicated fiber pair---for hardware synchronization and key distillation. For classical network integration, our setup employs a standard Gigabit Ethernet link, which multiplexes the Key Management System (KMS) traffic and the out-of-band management traffic for system configuration, continuous monitoring, and raw data extraction. Through this software stack, the platform enables real-time estimation of interference detection count, visibility, Quantum Bit Error Rate (QBER), and secret key rate, among other parameters, after sifting, error correction, and privacy amplification.

\subsection{Network Topology and Optical Channels}\label{subsec:opticalChannels}

The experimental campaign was structured around a standard point-to-point (P2P) network topology, directly connecting Alice (transmitter) and Bob (receiver) nodes. This architecture was implemented across two separate single-mode fiber scenarios to evaluate performance under different conditions: an indoor channel composed of spooled fibers and an outdoor configuration based on a deployed underground fiber link.

The indoor channel consisted of three fiber spools of standard single-mode fiber (ITU-T G.652.B) with nominal lengths of 20,199 m, 10,070 m, and 10,010 m. Two single-mode patch cords (Furukawa COA SM-MF 2001) of 2.60 m and 2.55 m connected the spools to the Alice and Bob QKD units, yielding a total channel length of 40,284 m. The fiber segments were coupled using standard FC/PC connectors via mating sleeves. The 20 km spool and one 10 km spool presented a nominal attenuation of 0.21 dB/km, while the remaining 10 km spool exhibited 0.19 dB/km. The actual total channel length and optical loss, characterized using a JDSU MTS-2000 Optical Time-Domain Reflectometer (OTDR) operating at 1550~nm, were measured as 40,343~m and 9.069~dB, respectively. A 3~dB optical attenuator was further incorporated into the link, resulting in an overall channel attenuation of 12.069~dB.

For the outdoor channel, we utilized a link from the Hermes Quantum Network. The link consists of an underground optical fiber cable connecting IME to CBPF, both located in the Urca neighborhood. While the straight-line distance between the institutions is approximately 804 m, the installed fiber path length is approximately 1.5 km. The cable, manufactured by FICAP, contains twelve standard single-mode fibers and features a jelly-filled core (thixotropic gel) and an Aluminum Polyethylene Laminate (APL) barrier for dual moisture protection. To co-locate both QKD units at the IME Photonics Laboratory, a loopback was performed at the CBPF node by connecting two fibers using optical connectors. Characterization using the same OTDR at 1550~nm yielded a measured total distance of 3,490~m and a fiber loss of 2.649~dB. Incorporating an additional 3~dB optical attenuator, the overall channel attenuation reached 5.649~dB. Table \ref{table:channels} summarizes these results.
\begin{table}[!htbp]
\centering
\caption{Summary of Quantum Channel Characteristics.}
\label{table:channels}
\begin{tabular}{lcc}
\toprule
\textbf{Parameter} & \textbf{Indoor} & \textbf{Outdoor} \\ \midrule
Fiber Length (m)   & 40,343                    & 3,490                       \\
Total Loss (dB)    & 12.07                      & 5.65                        \\
Configuration      & Spooled fiber             & Underground loop-back       \\ \bottomrule
\end{tabular}
\end{table}

In both indoor and outdoor configurations, the bidirectional service channel was routed over a dedicated pair of parallel fibers. For the indoor setup, this connection was established using the default 2-meter optical duplex patch cord supplied by the manufacturer, which consists of a standard single-mode fiber pair terminated with LC/PC connectors. For the outdoor configuration, the service channel utilized an identical loop-back configuration over two separate adjacent fibers within the same 12-fiber cable. In both cases, a 15~dB attenuator was incorporated into the service channel to keep the total attenuation within the manufacturer's required range.

\section{Experimental Procedure}\label{sec:experimentalProcedure}
\begin{figure*}[!t]
    \centering
    \resizebox{\textwidth}{!}{%
        \begin{tikzpicture}[x=0.21cm, y=1cm, >=Latex, font=\sffamily]
            \draw[->, thick, black!70] (-2,0) -- (85,0) node[right] {\footnotesize Time};
            \foreach \x/\label in {0/{23 Oct}, 25/{17 Nov}, 62/{24 Dec}, 66/{28 Dec}, 68/{30 Dec}, 71/{02 Jan}} {
                \draw[black!50] (\x, 0.1) -- (\x, -0.2) node[below, rotate=45, anchor=north east, font=\scriptsize] {\label};
            }
            \foreach \x/\label in {27/{19 Nov}, 55/{17 Dec}, 72/{03 Jan}, 78/{09 Jan}} {
                \draw[black!50] (\x, -0.1) -- (\x, 0.2) node[above, rotate=45, anchor=south west, font=\scriptsize] {\label};
            }
            \node[font=\bfseries\small, black!40] at (35, 2) {2025};
            \node[font=\bfseries\small, black!40] at (77, 2) {2026};
            \node[anchor=south west, font=\bfseries\small, black!70] at (0, 2.5) {PHASE I: Long-Term Analysis};
            \filldraw[fill=myCyanFill, draw=myCyanDraw, thick, rounded corners=2pt] (0, 0.5) rectangle (25, 1.3);
            \node[font=\scriptsize, align=center, text=myCyanDraw] at (12.5, 0.9) {Unregulated};
            \filldraw[fill=myOrangeFill, draw=myOrangeDraw, thick, rounded corners=2pt] (27, -0.5) rectangle (55, -1.3);
            \node[font=\scriptsize, align=center, text=myOrangeDraw] at (41, -0.9) {Unregulated};
            \node[anchor=south west, font=\bfseries\small, black!70] at (60, 2.5) {PHASE II: Controlled};
            \draw[dashed, black!30] (58, -2.5) -- (58, 2.5);
            \filldraw[fill=myCyanFill, draw=myCyanDraw, thick, rounded corners=2pt] (62, 0.5) rectangle (68, 1.3);
            \node[font=\scriptsize, text=myCyanDraw] at (65, 0.9) {Temperature};
            \filldraw[fill=white, draw=myCyanDraw, thick, dashed, rounded corners=2pt] (66, 1.4) rectangle (71, 1.9);
            \node[font=\scriptsize, text=myCyanDraw] at (68.5, 1.65) {Light};
            \filldraw[fill=myOrangeFill, draw=myOrangeDraw, thick, rounded corners=2pt] (72, -0.5) rectangle (78, -1.3);
            \node[font=\scriptsize, align=center, text=myOrangeDraw] at (75, -0.9) {Temperature};
            \node[anchor=east, font=\bfseries\scriptsize, text=myCyanDraw,align = center] at (-2, 0.9) {INDOOR \\ ($\sim$ 40 km) };
            \node[anchor=east, font=\bfseries\scriptsize, text=myOrangeDraw,align = center] at (-2, -0.9) {OUTDOOR \\ ($\sim$3.49km)};
        \end{tikzpicture}%
    }
    \caption{Timeline of the experimental campaign. The upper timeline (blue) represents the indoor channel activities, whereas the lower timeline (orange) represents the outdoor channel. The labels within the blocks specify  the parameter actively controlled during each period (if any). The dashed block represents the luminosity analysis period, whose data were omitted due to the absence of effect on the performance metrics.}
    \label{fig:timeline}
\end{figure*}
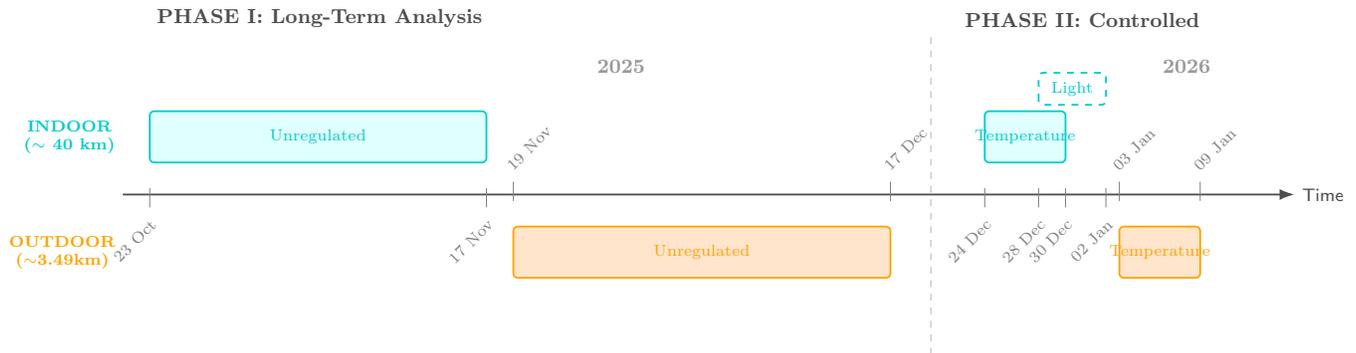

The analysis was divided into two main phases, detailed below. Fig.~\ref{fig:timeline} provides a schematic timeline of the complete experimental campaign. 

\subsection{Phase I: Long-Term Stability Analysis.}
To establish a baseline for the system's stability and robustness in real-world conditions, we performed continuous key generation over extended periods (approximately 1 month) for both channels. The indoor channel was monitored from October 23 to November 17, 2025. Subsequently, the outdoor channel underwent a similar long-run analysis from November 19 to December 18, 2025. During this phase, the system operated autonomously under standard, unregulated laboratory conditions, subject to routine human activity and ambient fluctuations. No strict control was imposed on the room parameters, with the sole exception of a background air conditioning unit, necessary to prevent the room temperature from exceeding the critical threshold for key generation. 
The use of this baseline cooling was strictly dictated by a temporary threshold configuration issue that halted key generation whenever Alice's internal temperature exceeded $40^\circ\text{C}$, which commonly corresponded to an ambient laboratory temperature of approximately $28^\circ\text{C}$. According to the manufacturer's specifications, the Clavis XGR is originally designed to operate normally at ambient temperatures ranging from $+5^\circ\text{C}$ to $+40^\circ\text{C}$. This configuration problem has since been resolved and did not compromise the validity of the data or the intrinsic functioning of the system, merely imposing an artificial thermal ceiling. The deployed outdoor optical fiber remained entirely unregulated and fully exposed to the high underground temperatures characteristic of the local climate, ensuring authentic environmental stress on the quantum channel.

\subsection{Phase II: Controlled Environmental Characterization.}
In the second phase, we aimed to isolate the effects of critical environmental variables---specifically temperature and ambient luminosity---on the QKD performance parameters.

Regarding temperature, we actively modulated the laboratory's internal temperature to assess the system's response under three distinct thermal regimes, achieved by adjusting the laboratory's air conditioning unit and restricting laboratory access strictly to necessary thermostat adjustments. Each temperature regime lasted approximately 2 days and was intended to establish an increasing temperature sequence.
The laboratory's ambient temperature and humidity were monitored using a DHT22 digital sensor interfaced with an Arduino microcontroller. To decouple thermal effects from optical noise, measurements were conducted under nominal darkness, with all lights kept switched off within the windowless laboratory. Data for the indoor channel were acquired between December 24 and 30, 2025, whereas the outdoor channel was measured between January 3 and 9, 2026.

From December 28, 2025, to January 2, 2026, to rule out optical noise as a confounding variable for the unjacketed indoor spools, we investigated the impact of ambient light on the indoor channel. The room temperature was kept as constant as possible to minimize thermal effects, and access to the laboratory was limited strictly to lighting adjustments. As expected, no statistical correlation with system performance was observed.

Throughout both experimental phases, data acquisition was continuously managed by the QNET central server---the platform's dedicated management software running on a local control server. Utilizing the management network established in our single-interface configuration, the system automatically recorded timestamped log files containing all real-time performance parameters. The primary metrics extracted for this analysis were the secret key rate, QBER, interferometric visibility, detection count, and the units' internal temperatures. All log files were exported via the Quantum Management System (QMS) Web Application and subsequently processed using custom Python scripts. 

\section{Results}\label{sec:results}

Given the high temporal resolution provided by the Clavis XGR system, plotting raw data over extended periods results in an excessive density of points that often obscures underlying trends. To address this, we applied temporal averaging to the dataset in most figures, enabling a clearer evaluation of the system's long-term evolution. Conversely, because temporal averaging inherently smooths out high-frequency events, we explicitly opted to present raw data in specific plots to highlight short-lived, anomalous estimations. Thus, raw data is displayed whenever capturing these brief transient disruptions is essential to the analysis. For the averaged plots, since the QNET central server does not provide uncertainty estimates, the standard deviation was adopted as the uncertainty measure, visualized as a shaded error band surrounding the mean curves, as usual.

Since temperature proved to be a dominant factor influencing parameter behavior, the QKD metrics are presented in most plots alongside the temperature profiles. The only exception is the key rate, which is plotted against visibility to emphasize the correlation between these two factors.

During Phase I, we relied exclusively on the internal temperature sensors of the QKD units. While these sensors provide a direct measure of the units temperatures, they record only hourly integer readings with a resolution of $1^\circ\text{C}$ (yielding an assumed uncertainty of $\pm 0.5^\circ\text{C}$). To obtain a more granular thermal profile, we recorded the external ambient temperature during Phase II using a dedicated sensor with a significantly higher sampling frequency and precision. 

As will be seen later in this section, the internal and external temperatures are strongly correlated; thus, utilizing the internal temperature for Phase I and the higher-resolution external temperature for Phase II does not compromise the analytical consistency. In fact, our quantitative analysis yielded a Spearman correlation coefficient of 0.96 between the internal and external temperatures, formally validating this approach. For consistency, the Spearman coefficient is adopted as the standard correlation metric throughout this paper.

To define the precise time windows for analysis, we considered the stability of both the system and the experimental conditions. First, we excluded the Clavis XGR initialization phase to filter out startup transients, as well as the thermal transition periods in Phase II before the target temperatures stabilized. Finally, we omitted data from periods in which Alice's internal temperature exceeded $40\,^{\circ}\mathrm{C}$, triggering the aforementioned artificial thermal ceiling that halted key generation. This thermal shutdown occurred twice: on December 17 at 06:35 UTC, ending the Phase I outdoor analysis, and on January 9 at 03:57 UTC, concluding the Phase II outdoor experiments.

Sections \ref{sec:longTermAnalysis} and \ref{sec:ControlledTemperature} detail the long-term stability analysis (Phase I) and the study under controlled environmental conditions (Phase II), respectively. In both analyses, we focused on parameters pivotal for QKD performance assessment: detection count, visibility, key rate, and QBER. 

\subsection{Long-Term Stability Analysis (Phase I)}\label{sec:longTermAnalysis}

This section details the system's performance during the long-term continuous runs (Phase I), conducted under unregulated laboratory conditions. In the absence of external weather logs for this phase, we contrast performance metrics with the internal temperature. As seen in Fig. \ref{fig:PhaseIindoorTemps},  both units undergo virtually identical thermal variations. Thus, we show Alice's temperature only, which was chosen because it is constantly higher than Bob's, facilitating the identification of periods in which the system operated near its critical thermal limits.

Table~\ref{table:PhaseImeanValues} presents the performance parameters averaged over the first experimental phase for both channels. Despite the lower detection rate of the spooled indoor channel relative to the outdoor channel, its overall performance was superior, characterized by a lower QBER and a higher key rate. This specific performance trend, addressed in Section~\ref{sec:discussion}, proved to be a consistent pattern across all experimental phases.
\begin{table}[htbp]
\centering
\caption{Mean Values and Standard Deviations for Indoor and Outdoor Scenarios throughout the first phase.}
\label{table:PhaseImeanValues}
\begin{tabular}{@{}lcc@{}}
\toprule
 & \textbf{Indoor} & \textbf{Outdoor} \\ 
                & \footnotesize \textbf{$\sim$ 40 km | 12 dB }& \footnotesize \textbf{$\sim$3.5 km | 6 dB} \\ 
                & \footnotesize 23 Oct -- 17 Nov& \footnotesize 19 Nov -- 17 Dec \\                
\midrule
Detection Count (Hz) & $60700 \pm 100$   & $64300 \pm 200$   \\
Visibility (\%)      & $98.2 \pm 0.6$  & $97.9 \pm 0.6$  \\
Key Rate (bps)       & $10500 \pm 500$   & $9400 \pm 500$    \\
QBER (\%)            & $0.7 \pm 0.2$   & $0.9 \pm 0.2$   \\
Temp. Alice (°C)     & $33 \pm 1$    & $34 \pm 1$    \\
Temp. Bob (°C)       & $30 \pm 1$    & $31 \pm 1$    \\
\bottomrule
\end{tabular}
\end{table}

\subsubsection{Indoor Channel (23 Oct - 17 Nov)}\label{sec:phaseIindoor}

\begin{figure}[!ht]
    \centering
    \includegraphics[width=.9\linewidth]{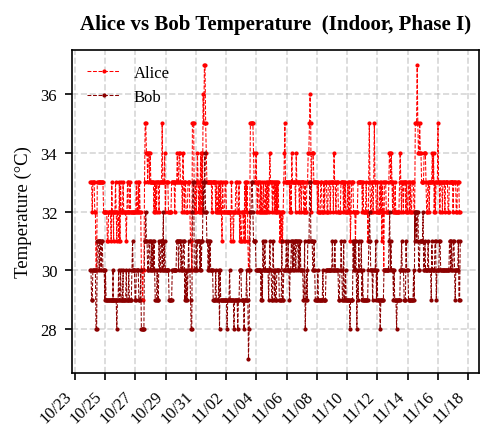}
    \caption{Internal temperatures of Alice and Bob Units during the long-term tests with fiber spools.}
    \label{fig:PhaseIindoorTemps}
\end{figure}

Despite the unregulated laboratory conditions, the Alice and Bob units maintained relatively stable temperatures, remaining mostly within a 3-degree variation band (Fig. \ref{fig:PhaseIindoorTemps}). Specifically, the average temperatures for Alice and Bob were $33 \pm 1\,^{\circ}\mathrm{C}$ and $30 \pm 1\,^{\circ}\mathrm{C}$, respectively, as shown in Table~\ref{table:PhaseImeanValues}. The highest temperatures recorded were $37\,^{\circ}\mathrm{C}$ for the Alice unit and $34\,^{\circ}\mathrm{C}$ for the Bob unit, both occurring on October 31. Subsequently, on November 7, the units experienced another thermal peak, with Alice and Bob recording $36\,^{\circ}\mathrm{C}$ and $33\,^{\circ}\mathrm{C}$, respectively. Finally, on November 14, Alice's temperature peaked once more at $37\,^{\circ}\mathrm{C}$, while Bob reached $32\,^{\circ}\mathrm{C}$.

\begin{figure}[!htbp]
    \centering
    \includegraphics[width=.9\columnwidth]{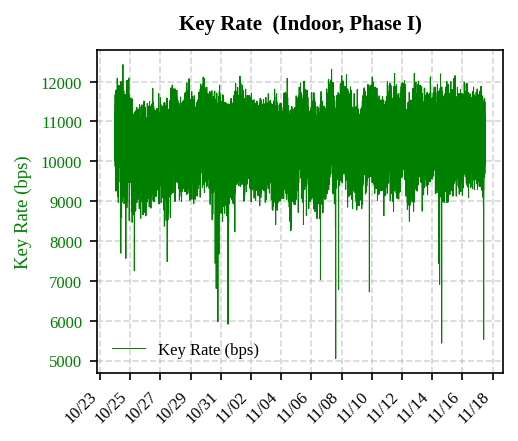}
    \caption{Key rate estimation during the indoor long-term performance test.}
    \label{fig:PhaseIindoorKeyRateRaw}
\end{figure}

\begin{figure}[!htbp]
    \centering
    \includegraphics[width=.85\columnwidth]{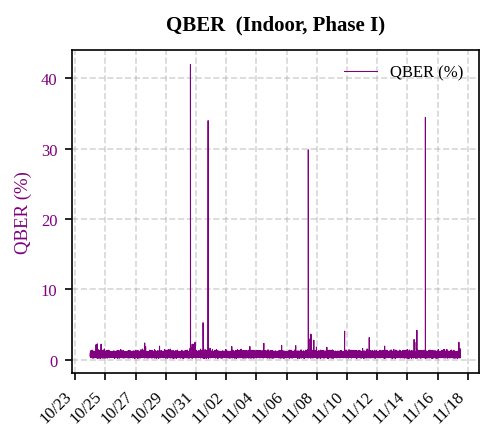}
    \caption{QBER during the indoor long-term performance test.}
    \label{fig:PhaseIindoorQBERRaw}
\end{figure}

\begin{figure}[!htbp]
        \centering
        \includegraphics[width=\columnwidth]{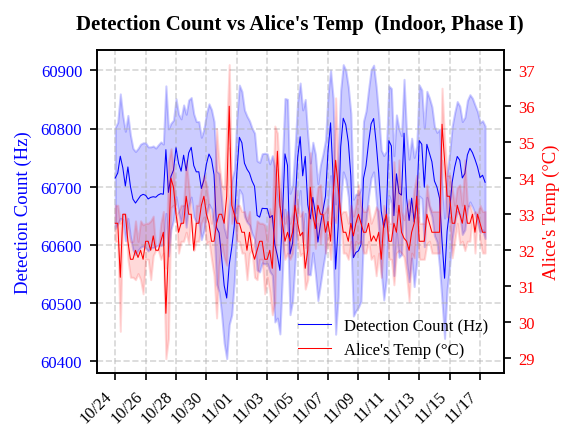}
        \caption{Detection count and Alice's temperature during the Phase I indoor experiment, averaged over 240 minutes.}
        \label{fig:PhaseIindoorCountVsTemp}
\end{figure}

The thermal peaks recorded on October 31, November 7, and November 14 were reflected in the system's performance metrics, aligning with the most pronounced deficiencies in both QBER and secret key rate estimations, as shown in Figs.\ref{fig:PhaseIindoorKeyRateRaw} and \ref{fig:PhaseIindoorQBERRaw}. The QBER exhibited virtually point-like anomalies, restricted to approximately 2-minute intervals, whereas the drop in the key rate was more notable. A similar pattern affects the detection count, as Fig. \ref{fig:PhaseIindoorCountVsTemp} shows.

\begin{figure}[!htbp]
        \centering
        \includegraphics[width=\columnwidth]{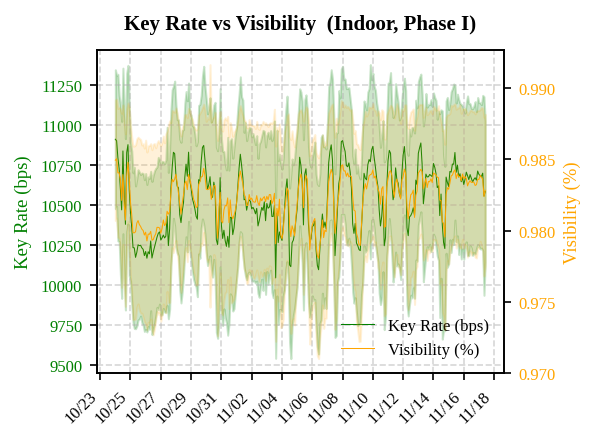}
        \caption{Key rate compared to visibility throughout the indoor long-term performance test, averaged over 120 min.}
        \label{fig:PhaseIindoorKeyVsVis}
\end{figure}

To analyze long-term trends and the correlation between visibility and key rate, we present both metrics averaged over 120-minute intervals. No anomalous estimations were observed for visibility, except for a spurious drop on October 30, which coincided with a period of substantial thermal instability. Fig. \ref{fig:PhaseIindoorKeyVsVis} illustrates the dependency of the key rate on the visibility, which presents a correlation coefficient of 0.85.

\subsubsection{Outdoor Channel (19 Nov - 17 Dec)}\label{sec:PhaseIoutdoor}
\begin{figure}[!htbp]
    \centering
    \includegraphics[width=\columnwidth]{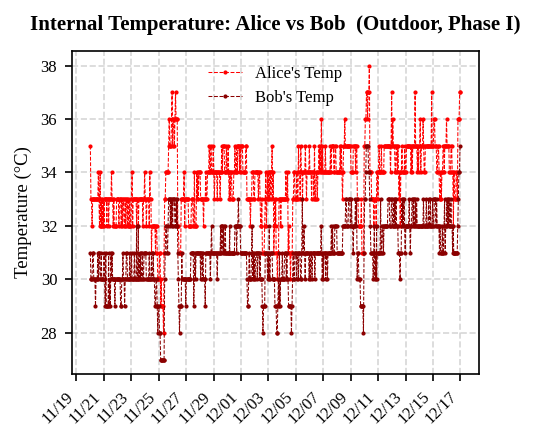}
    \caption{Alice and Bob internal temperatures throughout the long-term test with deployed fiber.}
    \label{fig:PhaseIOutdoorTemps}
\end{figure}

The temperature profiles for Alice and Bob are shown in Fig.~\ref{fig:PhaseIOutdoorTemps}. Whereas the unit temperatures remained relatively stable throughout the spooled indoor fiber tests, they exhibited considerable oscillations during this stage of the experiment.  As shown in Table~\ref{table:PhaseImeanValues}, average temperatures for both units were roughly 1°C higher than in the preceding stage, with Alice at 34°C and Bob at 31°C. As previously mentioned, the present analysis was concluded prior to the system's collapse, ensuring that the reported averages reflect the operational regime before the final failure.

\begin{figure}[!htbp]
    \centering
    \includegraphics[width=\linewidth]{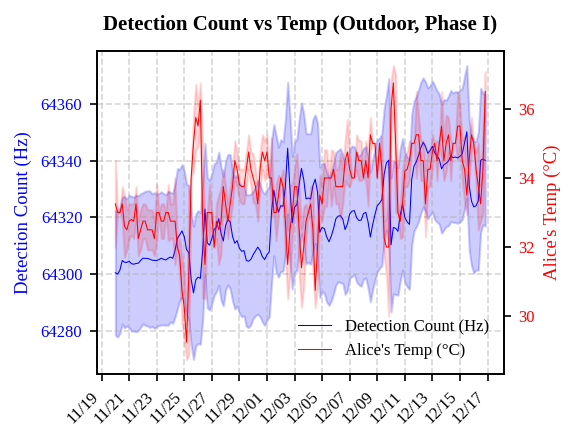}
    \caption{Detection count rates and Alice's internal temperature observed during the long-term outdoor fiber test, averaged over 240 minutes.}
    \label{fig:PhaseIOutdoorCountVsTemp}
\end{figure}

On December 10, at approximately 08:30 UTC, both Alice and Bob peaked at 38 °C and 35 °C, respectively. This maximum followed an 11-hour thermal increase from initial temperatures of 31 °C and 28 °C, configuring a relatively high thermal variation. A similar trend is observed from November 25 to 27. Between December 1 and 5, comparable oscillations occurred over a more extended period, exhibiting lower, yet still considerable, thermal gradients. These trends, as well as the relationship between detection counts and temperature, can be seen in Fig.~\ref{fig:PhaseIOutdoorCountVsTemp}.

Despite thermal variations being more pronounced at this stage, anomalous QBER, key rate, and visibility estimates were not observed. In what follows, we focus on the averaged trends to analyze the system's robustness.

\begin{figure}[!htbp]
    \centering
    \includegraphics[width=\columnwidth]{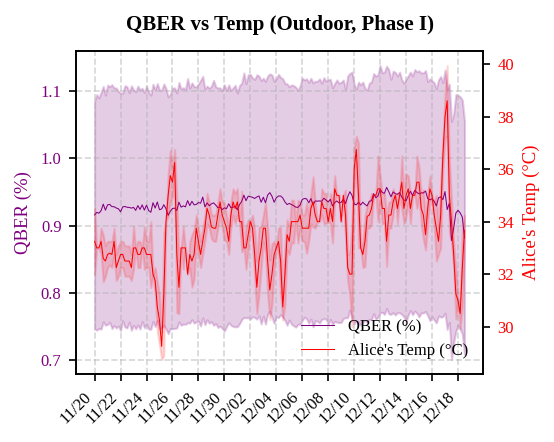}
    \caption{QBER and external temperature dynamics during the outdoor long-term test (averaged data).}
    \label{fig:PhaseIoutdoorQBERVsTemp}
\end{figure}

Figure \ref{fig:PhaseIoutdoorQBERVsTemp} contrasts the QBER evolution with the external temperature. The results indicate that the mean QBER remains remarkably robust against thermal fluctuations. As will be evidenced throughout this work, with the exception of virtually point-like anomalies, such as those of Section \ref{sec:phaseIindoor}, this stability constitutes a consistent pattern observed across all experimental stages.

\begin{figure}[!htbp]
    \centering
    \includegraphics[width=\linewidth]{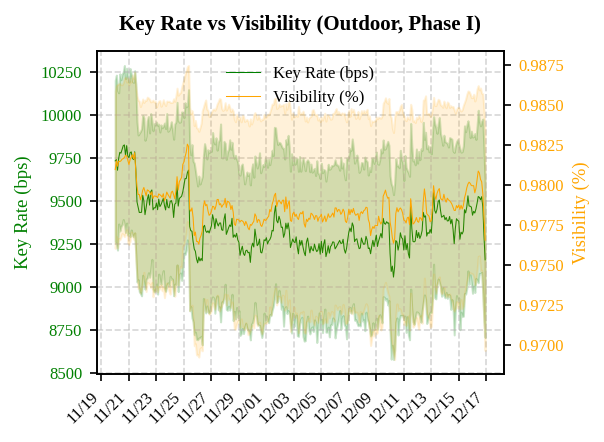}
    \caption{Key Rate and visibility observed during the outdoor long-term performance test, averaged over 240 minutes.}
    \label{fig:PhaseIOutdoorVisvsKey}
\end{figure}

Fig. \ref{fig:PhaseIOutdoorVisvsKey} displays the Key Rate plotted against visibility. Despite a visible correspondence between the two parameters, it is notably weaker than that observed in the indoor fiber scenario (Section \ref{sec:phaseIindoor}), yielding a correlation coefficient of 0.76.

\subsection{Controlled Temperature (Phase II)}\label{sec:ControlledTemperature}
This section details the system's performance during Phase II, covering the periods from December 24 to 30, 2025 (indoor channel), and January 3 to 9, 2026 (outdoor channel), in which we actively controlled the laboratory's temperature. As outlined in Section \ref{sec:experimentalProcedure}, three distinct thermal regimes were established for each channel, with each setting maintained for approximately 48 hours. 

Table~\ref{table:PhaseIImeanValues} presents performance metrics averaged over each regime for both channels. To ensure that the tabulated values reflect the system's behavior at the intended steady temperatures, data collected during thermal transitions were excluded from the calculation. Specifically, we discarded the interval starting 30 minutes prior to the adjustment of the air conditioning unit and extending to three hours after the change, allowing for thermal stabilization. These transition periods, though omitted from the Table, can be seen in the subsequent plots.
\begin{table}[!htbp]
\centering
\caption{Mean values and standard deviations for Indoor and Outdoor scenarios during the second experimental phase. The average external temperature for each thermal regime is indicated alongside the corresponding stage label.}
\label{table:PhaseIImeanValues}
\resizebox{\columnwidth}{!}{%
\addtolength{\tabcolsep}{-2pt}
\begin{tabular}{@{}lcc@{}}
\toprule
     & \textbf{Indoor} & \textbf{Outdoor} \\ 
                & \footnotesize \textbf{$\sim$ 40 km | 12 dB }& \footnotesize \textbf{$\sim$3.5 km | 6 dB} \\ 
                & \footnotesize 24 -- 30 Dec & \footnotesize 03 -- 09 Jan \\
\midrule
\textit{\textbf{Stage 1}} & $\mathbf{19.8 \pm 0.3^{\circ}\text{C}}$ & $\mathbf{20.3 \pm 0.2^{\circ}\text{C}}$ \\
\midrule
Det. Count (Hz)  & $60600 \pm 100$  & $64350 \pm 30$    \\
Visibility (\%)  & $98.3 \pm 0.5$ & $98.1 \pm 0.6$ \\
Key Rate (bps)   & $10500 \pm 500$  & $9400 \pm 500$    \\
QBER (\%)        & $0.7 \pm 0.2$  & $1.0 \pm 0.2$  \\
Temp. Alice (°C) & $32.1 \pm 0.6$   & $32.4 \pm 0.5$   \\
Temp. Bob (°C)   & $28.7 \pm 0.5$   & $28.9 \pm 0.8$   \\
\midrule
\textit{\textbf{Stage 2}} & $\mathbf{20.7 \pm 0.2^{\circ}\text{C}}$ & $\mathbf{20.0 \pm 0.2^{\circ}\text{C}}$ \\
\midrule
Det. Count (Hz)  & $60600 \pm 100$    & $64370 \pm 20$    \\
Visibility (\%)  & $98.5 \pm 0.4$ & $98.0 \pm 0.6$ \\
Key Rate (bps)   & $10700 \pm 500$  & $9300 \pm 500$    \\
QBER (\%)        & $0.7 \pm 0.2$  & $1.0 \pm 0.2$  \\
Temp. Alice (°C) & $32.8 \pm 0.5$   & $31.9 \pm 0.9$   \\
Temp. Bob (°C)   & $29.4 \pm 0.6$   & $28.3 \pm 0.5$   \\
\midrule
\textit{\textbf{Stage 3}} & $\mathbf{25.6 \pm 0.4^{\circ}\text{C}}$ & $\mathbf{26.5 \pm 1.0^{\circ}\text{C}}$ \\
\midrule
Det. Count (Hz)  & $60700 \pm 100$    & $64380 \pm 30$    \\
Visibility (\%)  & $98.2 \pm 0.6$ & $97.9 \pm 0.6$ \\
Key Rate (bps)   & $10500 \pm 500$  & $9400 \pm 500$    \\
QBER (\%)        & $0.7 \pm 0.2$  & $1.0 \pm 0.2$  \\
Temp. Alice (°C) & $36.9 \pm 0.6$   & $37.6 \pm 0.9$   \\
Temp. Bob (°C)   & $34.3 \pm 0.7$   & $34.6 \pm 0.9$   \\
\bottomrule
\end{tabular}%
}
\end{table}

Table \ref{table:PhaseIImeanValues} shows the same pattern observed in Table \ref{table:PhaseImeanValues}, with similar values and errors per channel for all parameters. Overall, the channel characteristics exerted a stronger influence on the system's baseline performance than the specific regimes under which the averages were calculated.

In this phase, we adopt the external laboratory temperature as the primary reference for our analysis, replacing Alice's internal temperature used in Phase I. This decision is driven by the superior data quality of the external sensors, which offer both a higher sampling frequency and greater precision. As shown in Figs. \ref{fig:PhaseIIindoorTemps} and \ref{fig:PhaseIIoutdoorTemps}, the internal and external temperatures remain tightly coupled, ensuring that this refinement in measurement precision enhances the analysis without compromising the consistency with previous results. 

Considering the shorter periods analyzed in this stage compared to Phase I, the averaging window for the system parameters was adjusted to 10 minutes. This adjustment allows for a finer temporal resolution, better suiting the analysis of the system's response to the controlled thermal steps.   

\subsubsection{Indoor Channel (24 - 30 Dec)}
\begin{figure}[!htbp]
    \centering
    \includegraphics[width=.9\linewidth]{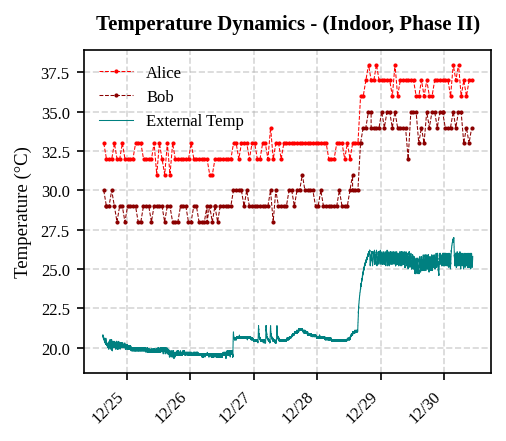} 
    \caption{Internal and external temperatures during the controlled temperature tests with spooled indoor fiber.}
    \label{fig:PhaseIIindoorTemps}
\end{figure}

In contrast to the unregulated fluctuations of the previous phase, the temperature profile in this stage exhibits clear, stepwise transitions corresponding to the three imposed regimes. Fig. \ref{fig:PhaseIIindoorTemps} demonstrates that, for the spooled indoor fiber, the internal temperatures of the units tightly track the external thermal trends. Specifically, the unit temperatures remain stable during the steady periods and increase in response to the external temperature steps. The correlation coefficient for the internal and external temperatures is 0.95. 

\begin{figure}[!htbp]
    \centering
    \includegraphics[width=\columnwidth]{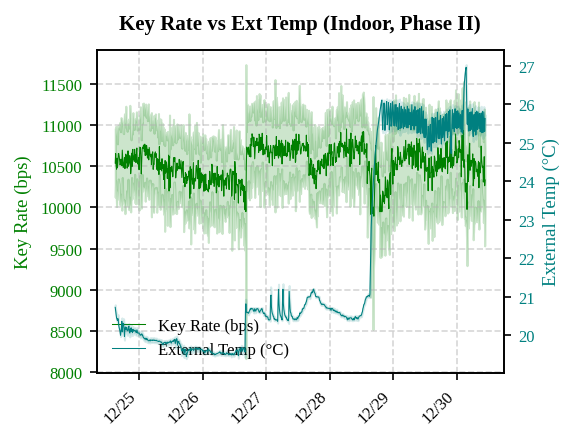}
    \caption{Key Rate evolution compared to the external laboratory temperature during Phase II. The curves represent data averaged over 10-minute intervals.}
    \label{fig:PhaseIIindoorKeyVsExtTemp}
\end{figure}

\begin{figure}[!htbp]
    \centering
    \includegraphics[width=.9\columnwidth]{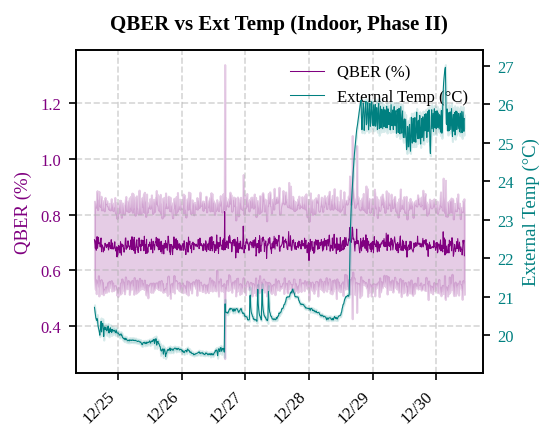}
    \caption{QBER and external temperature dynamics. Data points correspond to 10-minute averages.}
    \label{fig:PhaseIIindoorQBERVsExtTemp}
\end{figure}

Similar to the spooled indoor fiber tests during Phase I, anomalous estimations for both QBER and Key Rate were observed when the system underwent rapid temperature changes (transitions between stages, in this case). This can be seen in the averaged data indirectly, as accentuated standard deviations (error shadows). To visualize the general relationship between the system parameters and the external temperature, we show the data averaged over 10-minute intervals. Figs. \ref{fig:PhaseIIindoorKeyVsExtTemp} and \ref{fig:PhaseIIindoorQBERVsExtTemp} contrast the Key Rate and QBER with external temperature.

\begin{figure}[!htbp]
    \centering
    \includegraphics[width=.9\columnwidth]{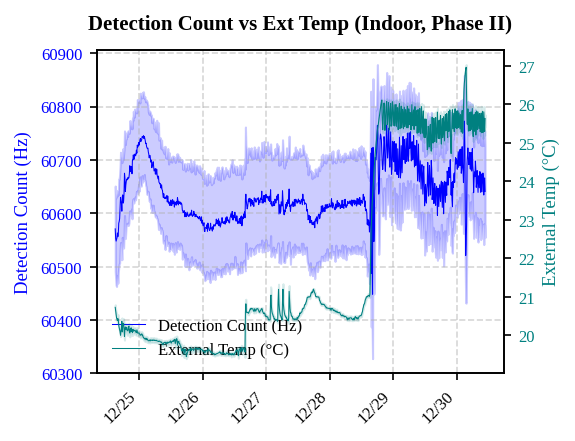}
    \caption{Detection count and external temperature evolution averaged over 30-minute intervals.}
    \label{fig:PhaseIIindoorCountVsExtTemp}
\end{figure}

The evolution of the detection count and its relationship with the external temperature are displayed in Fig. \ref{fig:PhaseIIindoorCountVsExtTemp}. We observe that the third stage, characterized by higher thermal instability compared to the previous ones, exhibited a higher variance in the detection count, as expected. Furthermore, the transition between Stages 2 and 3 ---marked by a significant temperature shift---was accompanied by a substantial relative oscillation in the detection frequency.

\begin{figure}[!htbp]
    \centering
    \includegraphics[width=1\columnwidth]{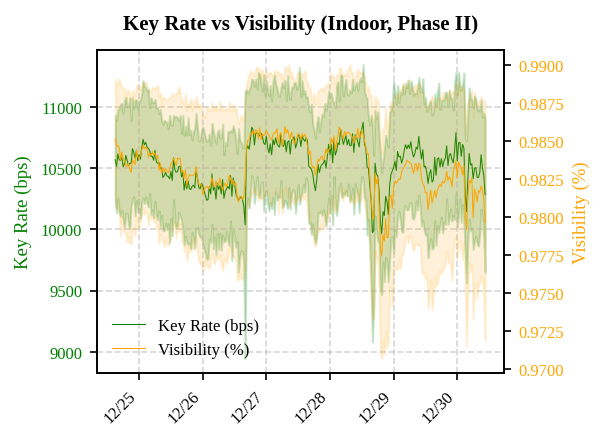}
    \caption{Correlation between Visibility and  Key Rate during Phase II. }
    \label{fig:PhaseIIindoorVisVsKeyrate}
\end{figure}

Finally, Fig. \ref{fig:PhaseIIindoorVisVsKeyrate} demonstrates a tight correlation between visibility and the key rate, reflecting the known interplay between optical alignment and performance. It presents a correlation coefficient of 0.91.

\subsubsection{Outdoor Channel (03 - 09 Jan )}
\begin{figure}[!htbp]
    \centering
    \includegraphics[width=.9\columnwidth]{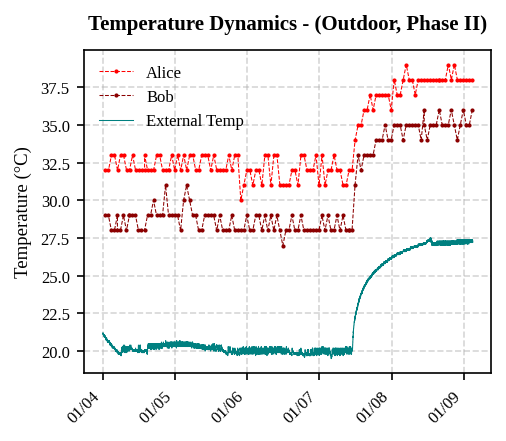}
    \caption{Internal and external temperatures during the controlled temperature tests with the installed fiber.}
    \label{fig:PhaseIIoutdoorTemps}
\end{figure}

Figure \ref{fig:PhaseIIoutdoorTemps} presents the raw temperature profiles for the installed fiber campaign. Due to external weather conditions influencing the laboratory environment, the mean temperature of Stage 1 was higher than that of Stage 2, deviating from the intended stepwise increase. Additionally, contrary to the trend observed in the indoor phase, thermal oscillations were more pronounced during the lower-temperature stages (1 and 2) than in the higher ones. It is also worth noting that the thermal gradient during the transition between Stages 2 and 3 was sharper than in the indoor test, with the average temperature of Stage 3 reaching approximately 1°C higher than its indoor counterpart. As in Phase I, the system experienced a collapse at the end of this high-temperature stage, which has been excluded from the current analysis. 

\begin{figure}[!htbp]
    \centering
    \includegraphics[width=\columnwidth]{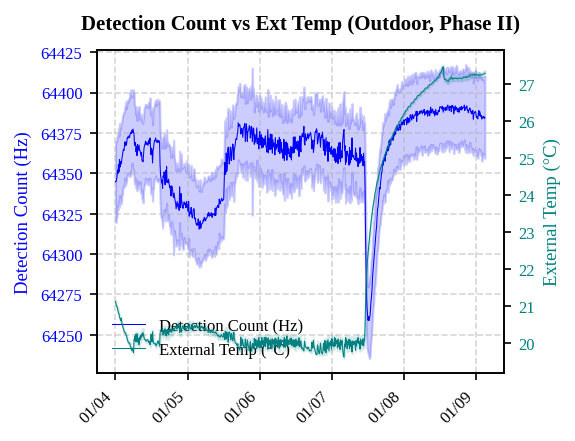}
    \caption{Detection count and external temperature evolution averaged over 10-minute intervals. }
    \label{fig:PhaseIIoutdoorCountVsExtTemp}
\end{figure}

The evolution of the detection count is shown in Fig. \ref{fig:PhaseIIoutdoorCountVsExtTemp}. Consistent with the indoor tests, periods characterized by higher thermal dispersion coincided with a larger variance in detection frequency. Moreover, a significant relative drop in the detection count was observed specifically during the sharp gradient transition between Stages 2 and 3. Note, however, that in absolute terms, this variation does not exceed that observed in the indoor fiber.

Despite the accentuated thermal gradient between Stages 2 and 3, no anomalous estimations for QBER or secret Key Rate were recorded during this phase. Thus, confirming the Phase I findings, such anomalous behavior was observed exclusively in the spooled fiber setup.

\begin{figure}[!htbp]
    \centering
    \includegraphics[width=\columnwidth]{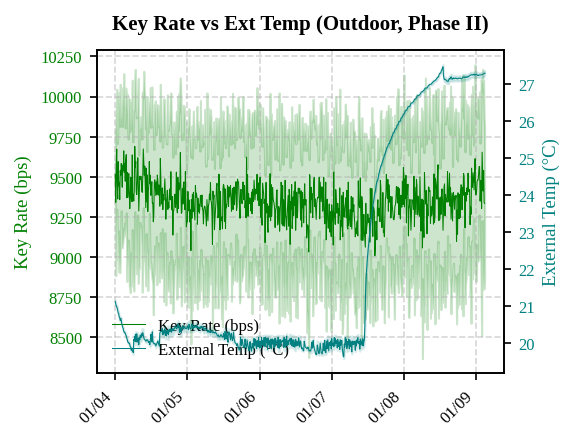}
    \caption{Key Rate evolution compared to the external laboratory temperature during phase II with installed fiber. The curves represent data averaged over 10-minute intervals.}
    \label{fig:PhaseIIoutdoorKeyVsExtTemp}
\end{figure}

In contrast to the indoor fiber results, the averaged key rate showed no notable variation during the temperature transitions, even in the presence of the steep thermal gradient. This stability, illustrated in Fig. \ref{fig:PhaseIIoutdoorKeyVsExtTemp}, is likely due to the absence of the aforementioned anomalies.

\begin{figure}[!htbp]
    \centering
    \includegraphics[width=.9\columnwidth]{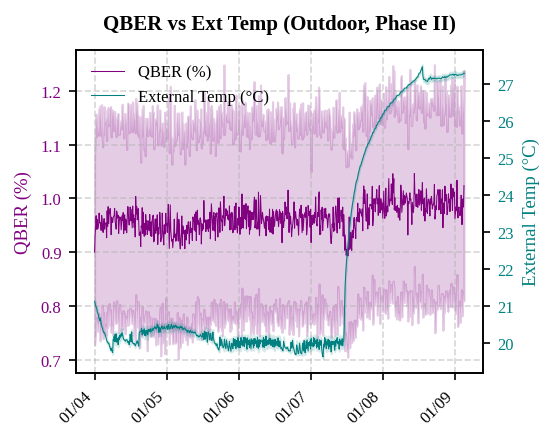}
    \caption{QBER and external temperature dynamics (10-minute averages).}
    \label{fig:PhaseIIoutdoorQBERVsExtTemp}
\end{figure}

The QBER, on the other hand, exhibited only a slight variation during the transition between Stages 2 and 3, as shown in Fig. \ref{fig:PhaseIIoutdoorQBERVsExtTemp}.

\begin{figure}[!htbp]
    \centering
    \includegraphics[width=\columnwidth]{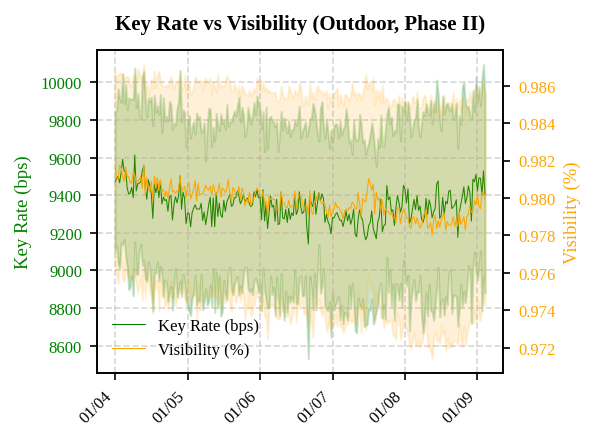}
    \caption{Correlation between Visibility and Key Rate during the outdoor phase (30-minute averages).}
    \label{fig:PhaseIIoutdoorVisVsKeyrate}
\end{figure}

Finally, Fig. \ref{fig:PhaseIIoutdoorVisVsKeyrate} compares the visibility with the secret key rate. The correlation remains significant, yet it exhibits higher dispersion compared to the indoor case---behavior that persisted across various averaging intervals analyzed---yielding a correlation coefficient of 0.86.

\section{Discussion}\label{sec:discussion}
The experimental results shown in this paper provide a comprehensive overview of the Clavis XGR system's behavior over long periods of time, under both normal and thermally controlled conditions. By evaluating the system across two fundamentally different optical links---namely a $40,343$~m indoor spooled fiber and a $3,490$~m outdoor deployed fiber---we assessed the impact of distinct channel characteristics on performance, while decoupling the environmental vulnerabilities of the quantum channel from those of the QKD hardware. The detailed evolution of performance parameters presented in this work serves as a useful reference for prospective users of the Clavis XGR, particularly in academic research, in which even minor system fluctuations are often highly relevant for experiments. 

In the indoor channel analysis, across both testing phases, we observed that sharp temperature gradients resulted in anomalous estimations of performance parameters---specifically key rate and QBER---over short time intervals (Figs. \ref{fig:PhaseIindoorKeyRateRaw}, \ref{fig:PhaseIindoorQBERRaw}, \ref{fig:PhaseIIindoorKeyVsExtTemp}, and \ref{fig:PhaseIIindoorQBERVsExtTemp}), whereas no such anomalies were observed in the outdoor tests. This discrepancy suggests that these transient disruptions are channel-related effects. Excluding these anomalous estimations, no significant performance degradation was observed due to pronounced thermal variations within the ambient temperature range of $19\,^{\circ}\mathrm{C}$ to $27\,^{\circ}\mathrm{C}$, provided the internal temperatures of the units remained below $40\,^{\circ}\mathrm{C}$---an artificial thermal ceiling imposed by a temporary configuration issue that has since been resolved. This demonstrates the overall robustness of the system against operational thermal fluctuations.

The correlation analysis extracted from our dataset provides quantitative insights into both the optical channel dynamics and the hardware's thermal susceptibility. In both phases, the correlation between visibility and secret key rate was found to be stronger in the indoor spooled fiber than in the deployed one. Specifically, in Phase I, we observed correlation coefficients of 0.85 and 0.76 for the indoor and outdoor channels, respectively, whereas Phase II yielded coefficients of 0.91 and 0.86. On the one hand, unlike the indoor spool, the real-world underground link is subjected to its own distinct thermal profile, localized mechanical stresses, micro-vibrations from urban traffic along the 3.5~km path. These confounding field variables introduce independent penalties to the error correction and privacy amplification budgets, partially decoupling the final secret key rate from the hardware's raw interferometric visibility. Additionally, the deployed outdoor channel exhibited a total attenuation of only 6~dB, which falls below the manufacturer's recommended minimum operational threshold of 10~dB for the quantum channel. Operating in this low-attenuation regime yields an excessive incident photon flux that potentially saturates the photodetectors and exacerbates afterpulsing effects,  restricting the secret key production and further degrading its dependence on the optical visibility. Finally, the near-perfect correlation ($r \approx 0.96$) between the laboratory's ambient temperature and the units' internal sensors demonstrates that the electronic components are tightly coupled to external environmental fluctuations, as expected.

Overall, the system demonstrated excellent resilience and operated in a high-efficiency regime across both channels. Visibility remained consistently high during the stability windows, fluctuating typically above $97\%$, indicating nearly ideal interferometric alignment. The QBER, in turn, remained below $1\%$ on average (approximately $0.70\%$ for the indoor scenario and $0.96\%$ for the outdoor scenario). The detection count rate was higher in the outdoor channel (approximately $64.3$~kHz compared to $60.6$~kHz for the indoor channel), an expected behavior given its considerably lower total attenuation. Meanwhile, the average key rate was higher in the indoor channel (typically $10.5$~kbps) than in the outdoor one (approximately $9.3$~kbps). This unbalance between detection count and key rate is consistent with both the higher intrinsic fluctuations of the deployed fiber, which entail discarding a larger fraction of bits during post-processing, and its below-threshold attenuation regime, which induces detector saturation and increases afterpulsing effects. Together, these effects penalize the final secure rate and outweigh the advantage of a higher volume of detected raw photons.

A notable aspect of this experimental campaign is its execution within a tropical climate. Most field deployments and characterizations of commercial QKD systems have historically been conducted in Europe, North America, and East Asia, regions whose climatic profiles generally differ from the persistent heat and environmental stress of the tropics. By operating the Clavis XGR within the Hermes Quantum Network, our setup exposed the hardware to the demanding thermal realities of a tropical deployment. While the indoor units required baseline cooling to prevent complete thermal shutdown, they nevertheless underwent significant thermal cycling characteristic of climatized server rooms in hot climates. Crucially, the outdoor deployed fiber was completely unregulated and continuously subjected to the elevated underground temperatures typical of the region. We hope that the data and insights provided in this study will be instrumental in guiding future research and infrastructure planning within the Hermes Quantum Network, as well as other emerging networks in Brazil, such as the Rio and Recife Quantum Networks \cite{temporao2024redeRio,Oliveira2025photonPairRecife}, and worldwide.

In sum, the comprehensive mapping of performance parameters presented in this work aims to clarify the practical operational dynamics of the QKD system, establishing a detailed baseline of the system's functionality, especially concerning external thermal conditions. Ultimately, the metrics and thermal thresholds identified herein provide scientists and general users with a clear understanding of the expected performance and operational boundaries of commercial QKD infrastructure across diverse deployment scenarios.

\section*{Acknowledgment}
The authors would like to thank FINEP for the financial support through the FINEP/MCTI Project No. 3310/24 – Research and Development of Quantum Technologies for National Security and Defense (Quantum II). This work has also been partially funded by the project EMBRAPII - QuIIN - CV-QKD na RRQ supported by Quantum Industrial Innovation (QuIIN) - EMBRAPII CIMATEC Competence Center in Quantum Technologies, with financial resources from the PPI IoT/Manufatura 4.0 of the MCTI grant number 053/2023, signed with EMBRAPII.

\bibliographystyle{apsrev4-2}
\bibliography{bibliography}

@inproceedings{temporao2024redeRio,
  author    = {Tempor{\~a}o, Guilherme P. and Melo, Fernando R. V. Bandeira de and Khoury, Antonio Z.},
  title     = {The Rio Quantum Network: a reconfigurable hybrid multi-user metropolitan quantum key distribution network},
  booktitle = {Anais do I Workshop de Redes Qu{\^a}nticas (WQuNets)},
  year      = {2024},
  pages     = {19--24},
  publisher = {Sociedade Brasileira de Computa{\c{c}}{\~a}o},
  address   = {Porto Alegre},
  doi       = {10.5753/wqunets.2024.2872},
  url       = {https://doi.org/10.5753/wqunets.2024.2872}
}

@article{bennett2014quantum,
	title={Quantum cryptography: Public key distribution and coin tossing},
	author={Bennett, Charles H and Brassard, Gilles},
	journal={Theoretical computer science},
	volume={560},
	pages={7--11},
	year={2014},
	publisher={Elsevier}
}

@article{brendel1999pulsed,
  title={Pulsed energy-time entangled twin-photon source for quantum communication},
  author={Brendel, J{\"u}rgen and Gisin, Nicolas and Tittel, Wolfgang and Zbinden, Hugo},
  journal={Physical review letters},
  volume={82},
  number={12},
  pages={2594},
  year={1999},
  publisher={APS}
}

@article{hwang2003quantum,
  title={Quantum key distribution with high loss: toward global secure communication},
  author={Hwang, Won-Young},
  journal={Physical review letters},
  volume={91},
  number={5},
  pages={057901},
  year={2003},
  publisher={APS}
}

@article{lo2005decoy,
  title={Decoy state quantum key distribution},
  author={Lo, Hoi-Kwong and Ma, Xiongfeng and Chen, Kai},
  journal={Physical review letters},
  volume={94},
  number={23},
  pages={230504},
  year={2005},
  publisher={APS}
}

@article{lo2005efficientQKDandProof,
author = {Lo, Hoi-Kwong and Chau, H. F. and Ardehali, M.},
title = {Efficient Quantum Key Distribution Scheme and a Proof of Its Unconditional Security},
year = {2005},
issue_date = {April 2005},
publisher = {Springer-Verlag},
address = {Berlin, Heidelberg},
volume = {18},
number = {2},
issn = {0933-2790},
url = {https://doi.org/10.1007/s00145-004-0142-y},
doi = {10.1007/s00145-004-0142-y},
journal = {J. Cryptol.},
month = apr,
pages = {133–165},
numpages = {33}
}

@article{tittel2000CryptographyEnergyTimeBell,
  title = {Quantum Cryptography Using Entangled Photons in Energy-Time Bell States},
  author = {Tittel, W. and Brendel, J. and Zbinden, H. and Gisin, N.},
  journal = {Phys. Rev. Lett.},
  volume = {84},
  issue = {20},
  pages = {4737--4740},
  numpages = {0},
  year = {2000},
  month = {May},
  publisher = {American Physical Society},
  doi = {10.1103/PhysRevLett.84.4737},
  url = {https://link.aps.org/doi/10.1103/PhysRevLett.84.4737}
}

@article{kawakami2025securityBB84passive,
  title = {Security of the BB84 protocol with passive biased basis choice by the receiver},
  author = {Kawakami, Shun and Taniguchi, Atsushi and Tonomura, Yoshihide and Takasugi, Koichi and Azuma, Koji},
  journal = {Phys. Rev. Appl.},
  volume = {24},
  issue = {5},
  pages = {054070},
  numpages = {19},
  year = {2025},
  month = {Nov},
  publisher = {American Physical Society},
  doi = {10.1103/2pzb-ssrs},
  url = {https://link.aps.org/doi/10.1103/2pzb-ssrs}
}

@manual{IDQ2025clavisBruchure,
  organization = {{ID Quantique}},
  title        = {Clavis XGR QKD System Brochure},
  year         = {2025},
  month        = {March},
  note         = {Copyright 2026 ID Quantique SA - All rights reserved - G.192.0121-PB-2.4 - Specifications as of February 2026. Accessed: April 15, 2026},
  url          = {https://www.idquantique.com/quantum-safe-security/products/clavis-xg-qkd-system/}
}

@article{Stanley2022ProgressQKDNetworks,
doi = {10.1088/1742-6596/2416/1/012001},
url = {https://doi.org/10.1088/1742-6596/2416/1/012001},
year = {2022},
month = {dec},
publisher = {IOP Publishing},
volume = {2416},
number = {1},
pages = {012001},
author = {Stanley, M and Gui, Y and Unnikrishnan, D and Hall, S.R.G and Fatadin, I},
title = {Recent Progress in Quantum Key Distribution Network Deployments and Standards},
journal = {Journal of Physics: Conference Series}
}

@article{Pirandola2020advancesQuantumCryptography,
author = {S. Pirandola and U. L. Andersen and L. Banchi and M. Berta and D. Bunandar and R. Colbeck and D. Englund and T. Gehring and C. Lupo and C. Ottaviani and J. L. Pereira and M. Razavi and J. Shamsul Shaari and M. Tomamichel and V. C. Usenko and G. Vallone and P. Villoresi and P. Wallden},
journal = {Adv. Opt. Photon.},
number = {4},
pages = {1012--1236},
publisher = {Optica Publishing Group},
title = {Advances in quantum cryptography},
volume = {12},
month = {Dec},
year = {2020},
url = {https://opg.optica.org/aop/abstract.cfm?URI=aop-12-4-1012},
doi = {10.1364/AOP.361502},
}

@article{Zhang2025towardsGlobalQKD,
  author    = {Zhang, Haoran and Zhu, Haotao and He, Ruihua and Zhang, Yan and Ding, Chao and Hanzo, Lajos and Gao, Weibo},
  title     = {Towards global quantum key distribution},
  journal   = {Nature Reviews Electrical Engineering},
  year      = {2025},
  month     = {dec},
  volume    = {2},
  number    = {12},
  pages     = {806--818},
  doi       = {10.1038/s44287-025-00238-7},
  url       = {https://doi.org/10.1038/s44287-025-00238-7},
  issn      = {2948-1201}
}

@misc{rehman2026companies,
  author       = {Rehman, Mohib Ur},
  title        = {Quantum-Safe Cryptography: Companies Across the Landscape -- 2026},
  howpublished = {The Quantum Insider},
  year         = {2026},
  month        = {apr},
  url          = {https://thequantuminsider.com/2026/03/25/25-companies-building-the-quantum-cryptography-communications-markets/},
  urldate      = {2026-04-16},
  note         = {Accessed on: 16 Apr. 2026}
}

@article{bozzio2025beyondQKD,
  title = {Quantum cryptography beyond key distribution: Theory and experiment},
  author = {Bozzio, Mathieu and Cr\'epeau, Claude and Wallden, Petros and Walther, Philip},
  journal = {Rev. Mod. Phys.},
  volume = {97},
  issue = {4},
  pages = {045006},
  numpages = {53},
  year = {2025},
  month = {Dec},
  publisher = {American Physical Society},
  doi = {10.1103/p84v-1xqv},
  url = {https://link.aps.org/doi/10.1103/p84v-1xqv}
}

@article{hofbauer2018temperatureDarkCountSinglePhoton,
author = {Hofbauer, Michael and Steindl, Bernhard and Zimmermann, Horst},
title = {Temperature Dependence of Dark Count Rate and After Pulsing of a Single-Photon Avalanche Diode with an Integrated Active Quenching Circuit in 0.35 μm CMOS},
journal = {Journal of Sensors},
volume = {2018},
number = {1},
pages = {9585931},
doi = {https://doi.org/10.1155/2018/9585931},
url = {https://onlinelibrary.wiley.com/doi/abs/10.1155/2018/9585931},
eprint = {https://onlinelibrary.wiley.com/doi/pdf/10.1155/2018/9585931},
year = {2018}
}

@inproceedings{ma2021thermalBirefringence,
author = {Honghao Ma and Danni Liu and Changkun Feng and Lishuang Feng},
title = {{Thermal sensitivity of the birefringence of photonic-bandgap fiber}},
volume = {12066},
booktitle = {AOPC 2021: Micro-optics and MOEMS},
editor = {Yuelin Wang and Huikai Xie and Yun-Feng Xiao},
organization = {International Society for Optics and Photonics},
publisher = {SPIE},
pages = {1206614},
year = {2021},
doi = {10.1117/12.2606185},
URL = {https://doi.org/10.1117/12.2606185}
}

@book{Agrawal2010fiberOpticSystems,
  author    = {Agrawal, Govind P.},
  title     = {Fiber-Optic Communication Systems},
  publisher = {John Wiley \& Sons, Ltd},
  year      = {2010},
  isbn      = {9780470918524},
  doi       = {10.1002/9780470918524},
  url       = {https://onlinelibrary.wiley.com/doi/book/10.1002/9780470918524}
}

@book{Sze2006semiconductorDevices,
  author    = {Sze, Simon M. and Ng, Kwok K.},
  title     = {Physics of Semiconductor Devices},
  publisher = {John Wiley \& Sons, Ltd},
  year      = {2006},
  isbn      = {9780470068328},
  doi       = {10.1002/9780470068328},
  url       = {https://onlinelibrary.wiley.com/doi/book/10.1002/9780470068328}
}

@article{Hadfield2009singlePhotonDetectors,
  author    = {Hadfield, Robert H.},
  title     = {Single-photon detectors for optical quantum information applications},
  journal   = {Nature Photonics},
  year      = {2009},
  month     = dec,
  volume    = {3},
  number    = {12},
  pages     = {696--705},
  doi       = {10.1038/nphoton.2009.230},
  url       = {https://doi.org/10.1038/nphoton.2009.230},
  issn      = {1749-4893}
}

@article{moreira2024nationalSovereignty,
  title={Quantum technologies: a matter of national sovereignty},
  author={Moreira, Fernando Manuel Ara{\'u}jo and Carneiro, V{\'i}tor Gouv{\^e}a Andrezo and Galdino, Juraci Ferreira},
  journal={Revista da Escola de Guerra Naval},
  volume={30},
  number={3},
  pages={643--686},
  year={2024},
  doi={10.21544/2359-3075.30330},
  url={https://portaldeperiodicos.marinha.mil.br/index.php/revistadaegn/article/view/5909}
}

@inproceedings{zanetti2025brazilianQKDspace,
author = {Zanetti, Marcelo and De Ávila, Leandro and Carneiro, Vítor},
year = {2026},
month = {01},
pages = {},
booktitle = {III Congresso Aeroespacial Brasileiro},
title = {Conceptual study and requirements for a brazilian QKD space mission},
doi = {10.29327/9786527220794.1443143}
}

@article{Oliveira2025photonPairRecife,
  title = {Photon-Pair Source for the Recife Quantum Network},
  author = {Oliveira, Nallyson W. S. and Siqueira, Andr\'{e} C. A. and Cris\'{o}stomo, Gabriel A. and Martins, Paulo and Freitas, Leandro L. L. and de Almeida, Alexandre A. C. and Acioli, L\'{u}cio H. and Martins-Filho, Joaquim F. and Coelho, Leonardo D. and Ferraz, Jos\'{e} and Felinto, Daniel},
  journal = {Brazilian Journal of Physics},
  volume = {56},
  number = {2},
  pages = {58},
  year = {2025},
  doi = {10.1007/s13538-025-01971-y},
  url = {https://doi.org/10.1007/s13538-025-01971-y}
}

\end{document}